# Identification of Nuclei Exhibiting the *SU*(3) Dynamical Symmetry

**A. Y. Abul-Magd, S. A. Mazen, M. Abdel-Mageed, and A. Al-Sayed***
*Faculty of Science, Zagazig University, Zagazig, Egypt.*

## ABSTRACT

**We consider the possibility of identifying nuclei exhibiting the *SU*(3) dynamical symmetry as those having excitation energy ratio $R_{4/2} \geq 3.25$. For this purpose, we consider the level statistics of some of these nuclei and perform interacting boson model (IBM) calculation of level schemes, and electromagnetic transition rates. We show that only some of these nuclei may be considered as good examples of the *SU*(3) dynamical symmetry.**

***Key Words: Nuclear Structure/ Rotational nuclei/ nuclear chaos.***

## 1-Introduction

Since Bohr and Mottelson proposed the geometrical model for nuclear collective motion, the search for nuclei suitable for collective description never stopped [1]. Several studies in this direction involve the interacting boson model (IBM) by Arima and Iachello [2]. The simplest formulation of the model (IBM-I) describes the even-even nucleus as an inert core combined with $N$ bosons which represent pairs of identical nucleons. Each boson has two states: an $L = 0$ ($s$-boson) state and an $L = 2$ ($d$-boson) state with five orientations. Thus, the Hilbert space of the IBM carries an irreducible representation of the group $U(6)$ which has several subgroups. Three dynamical symmetries are obtained by constructing the chains of subgroups of the $U(6)$ group that end with the angular momentum group $SO(3)$, since nuclear states have good angular momentum. The corresponding dynamical symmetries are usually referred to as the $U(5)$, $SU(3)$ and $O(6)$ limits. Different phases of the model can be associated to these dynamical symmetries. A spherical nucleus is related to the $U(5)$ symmetry, a well deformed nucleus is related to the $SU(3)$ symmetry, while γ-unstable nuclei correspond to the $O(6)$ symmetry. Systematic studies of medium and heavy nuclei [3] led to the identification of a small number of nuclei belonging to each of the three dynamical symmetries. The transitional regions between the three phases have also been a subject of extensive investigations (e.g. [4, 5]).

A weak point of the IBM analysis is how to define an "order parameter" as a probe of the variety of nuclear structure. Several possibilities have been considered. Among these is the ratio $R_{4/2}$ of excitation energies of the first $4^+$ and the first $2^+$ excited states. The IBM calculation of energy levels yields values of $R_{4/2}$ = 2.00, 3.33, and 2.50 for the dynamical symmetries $U(5)$, $SU(3)$, and $O(6)$, respectively. In this respect, a systematic analysis of the nearest neighbor-spacing distributions for $2^+$ levels of even-even nuclei [6]. The chaoticity parameter $f$ for nuclei obtained in this analysis has deep minima at these values of $R_{4/2}$. Such regularity is interpreted as an evidence for the enhanced correlation between the energy levels of nuclei having certain symmetry.

=====================================================

*E-mail: a.alsayed@zu.edu.eg

The $SU(3)$ group was recognized in the late 1950's to be reflective of the symmetries inherent in the properties of the elementary hadrons [7,8]. Later, Elliot used the SU(3) in the classification of rotational states in non-spherical nuclei [9]. Elliott's model played an essential role in the theory of nuclear structure [10-12]. It was visualized as a connecting bridge between the shell and collective models [13]. The importance and development of the $SU(3)$ model is reviewed in Ref [14]. It is clearly of interest to search for nuclei that display more closely the $SU(3)$ limit of the IBM. In this context several studies (e.g.[15-25]) are performed.

The aim of the present work is to test the probability of using the $R_{4/2}$ ratio to identify nuclei which can be described as good example of the $SU(3)$ dynamical symmetry.

## 2-Data Set

The present analysis involves data on low-lying levels of selected even-even nuclei with spins ranging from 0 to 6, which are taken from the Nuclear Data Sheets until July 2009. We considered nuclei in which the spin-parity $J^{\pi}$ assignments of at least five consecutive levels are definite. When the spin-parity assignment is uncertain and while the most probable value appears in brackets, we accept this value. We terminate the sequence in each nucleus when we reach a level with unassigned $J^{\pi}$, or when an ambiguous assignment involved a spin-parity among several possibilities, as e.g. $J^{\pi} = (2^{+}, 4^{+})$. We make an exception when only one such level is followed by several definitely assigned levels containing at least two levels of the same spin-parity, provided that this ambiguous level is found in a similar position in the spectrum of a neighboring nucleus. However, this situation has occurred for less than 5% of the levels considered. In this way, we have obtained 38 nuclei with definite $R_{4/2} \geq 3.25$ ratio given in table (1), 18 of them having at least five consecutive $2^{+}$ levels.

## 3-Attestation of Rotational Nuclei

In this section, we begin by testing the collective behavior of nuclei in the range $R_{4/2} \geq 3.25$ in terms of their level spacing distribution and level cross correlation. Then a detailed analysis of each nucleus is introduced in subsection 3-3.

### 3-1:-Nearest Neighbor Spacing Distribution

A nucleus consists of a large number of nucleons which interact via complicated forces. It is an extremely nonlinear system, which is most probably fully chaotic. The presence of a dynamical symmetry in the nucleus implies the conservation of one or more quantum numbers other than energy, angular momentum, and parity. In the case of $SU(3)$, this will be the $K$ quantum number. The energy-level spectrum (with fixed spin and parity) will then consist of superposition of nearly independent sequences, one for each value of the presumably conserved quantum numbers. Let $f_i$ to be the fractional level density of the sequence $i$. The nearest neighbor spacing (NNS) distribution of the total spectrum is approximately given by

$$P(s,f) = [1-f + q(f)\frac{\pi s}{2}]\exp[-(1-f)s - q(f)\frac{\pi s^{2}}{4}] \tag{1}$$

where $f = \sum_{i=1}^{m} f_i^{2}$ is the mean fractional level density for the superimposed sequences and the single parameter characterizing the distribution. The derivation of this formula is given in [6, 26]. The parameter $q$ is determined by the condition of unit mean spacing, and approximated by

$$q(f) = f\,(0.7 + 0.3f)\,. \qquad (2)$$

For a superposition of a large number of $m$ sequences, $f$ is of the order $1/m$. In the limit of $m \to \infty$, $f \to 0$. This produces a Poisson distribution. On the other hand, if $f \to 1$, NNS distribution approaches the Wigner distribution. Therefore, $f$ is referred to as the chaoticity parameter. A value of $f \approx 0$ obtained in the analysis of a nuclear spectrum does not necessarily indicate an actually regular behavior, but may also indicate the presence of a certain symmetry that divides the spectrum into independent level sequences. We believe that this is the case with collective nuclei.

We shall now consider NNS distribution of the levels of nuclei selected by criteria mentioned in section 2. In a previous investigation [6], two intervals of $R_{4/2}$ were considered to probe collective behavior, namely $3.20 \le R_{4/2} < 3.30$ and $R_{4/2} \ge 3.30$, corresponding to 113, and 79 spacings respectively. In current study we consider only 18 nuclei of the 58 ones that fall in the range $R_{4/2} \ge 3.25$. Comparing with the previous investigation, we arrive to two main conclusions. First, the general trend of collective behavior found in both investigations is conserved even when a fewer number of nuclei are involved, as in the current study. Second, seven years separate the data collected in Ref. [6] from the ones in the current study. In spite of this, the values of $f$ obtained in both studies are nearly the same. We hope that this result may remain valid also in the forthcoming years.

The following nuclei had been considered in the presence analysis: $^{174}$Hf, $^{160}$Dy, $^{230}$Th, $^{184}$W, $^{232}$Th, $^{182}$W, $^{232}$U, $^{178}$Hf, $^{170}$Yb, $^{162}$Dy, $^{234}$U, $^{164}$Dy, $^{172}$Yb, $^{240}$Pu, $^{168}$Er, $^{170}$Er, $^{246}$Cm, and $^{250}$Cf. The result of the analysis for NNS distribution is given in Fig. 1. The best-fit value of the chaoticity parameter $f$ is $0.58 \pm 0.10$. This may indicate the presence of an approximately conserved quantum number that takes at least two values in the investigated spectra. The $K$ quantum number of the $SU(3)$ group is a good candidate.

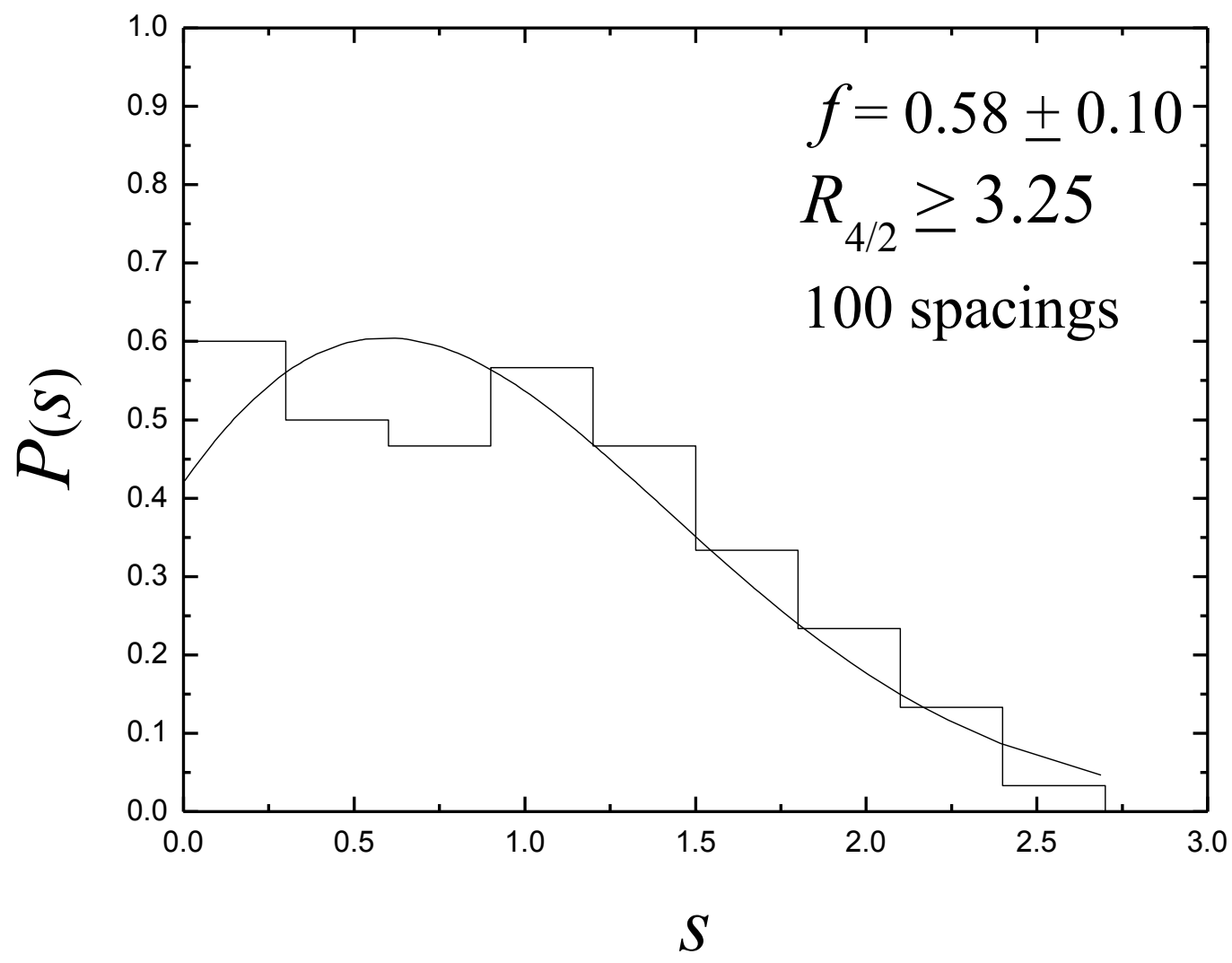

Fig.(1): The NNS distribution for 18 selected nuclei having at least five consecutive $2^+$ levels in the range of $R_{4/2} \ge 3.25$.

## 3-2: Level Cross-Correlation

A Cross-correlation coefficient between two sets of variables determines the extent to which these variables are linearly related. We here calculate the cross-correlation coefficients for each pair of energy levels that belong to nuclei having $R_{4/2} \geq 3.25$. The cross-correlation coefficient for two levels labeled by *i* and *j* is defined as [27]

$$C_{ij} = \frac{\sum_{n=1}^{N} \left[ E_i(n) - \bar{E}_i \right] \left[ E_j(n) - \bar{E}_j \right]}{\sqrt{\sum_{n=1}^{N} \left[ E_i(n) - \bar{E}_i \right]^2} \sqrt{\sum_{n=1}^{N} \left[ E_j(n) - \bar{E}_j \right]^2}}, \tag{3}$$

where $\bar{E}_i$ and $\bar{E}_j$ are the means of the corresponding series of energy levels, and $N = \min(N_i, N_j)$. We then calculate the mean value $\langle C \rangle$ of the coefficients of each of these levels $\langle C \rangle_i = \Sigma_i C_{ij}$. We have considered the first excited states with spins $J = 0 - 6$ including only positive parity which falls with the calculations of IBM. Obtaining a value of $\langle C \rangle$ close to 1 suggests that the corresponding nuclei belonging to the group, at least in the state of label *i*, belongs to the same collective state. The results of calculations are shown in Fig. (2). Most of the mean values $\langle C \rangle$ are close to 1 for nuclei belonging to the investigated interval. In average $\langle C \rangle = 0.94 \pm 0.06$, suggesting that all (or the majority) of nuclei in this interval having a common collective behavior.

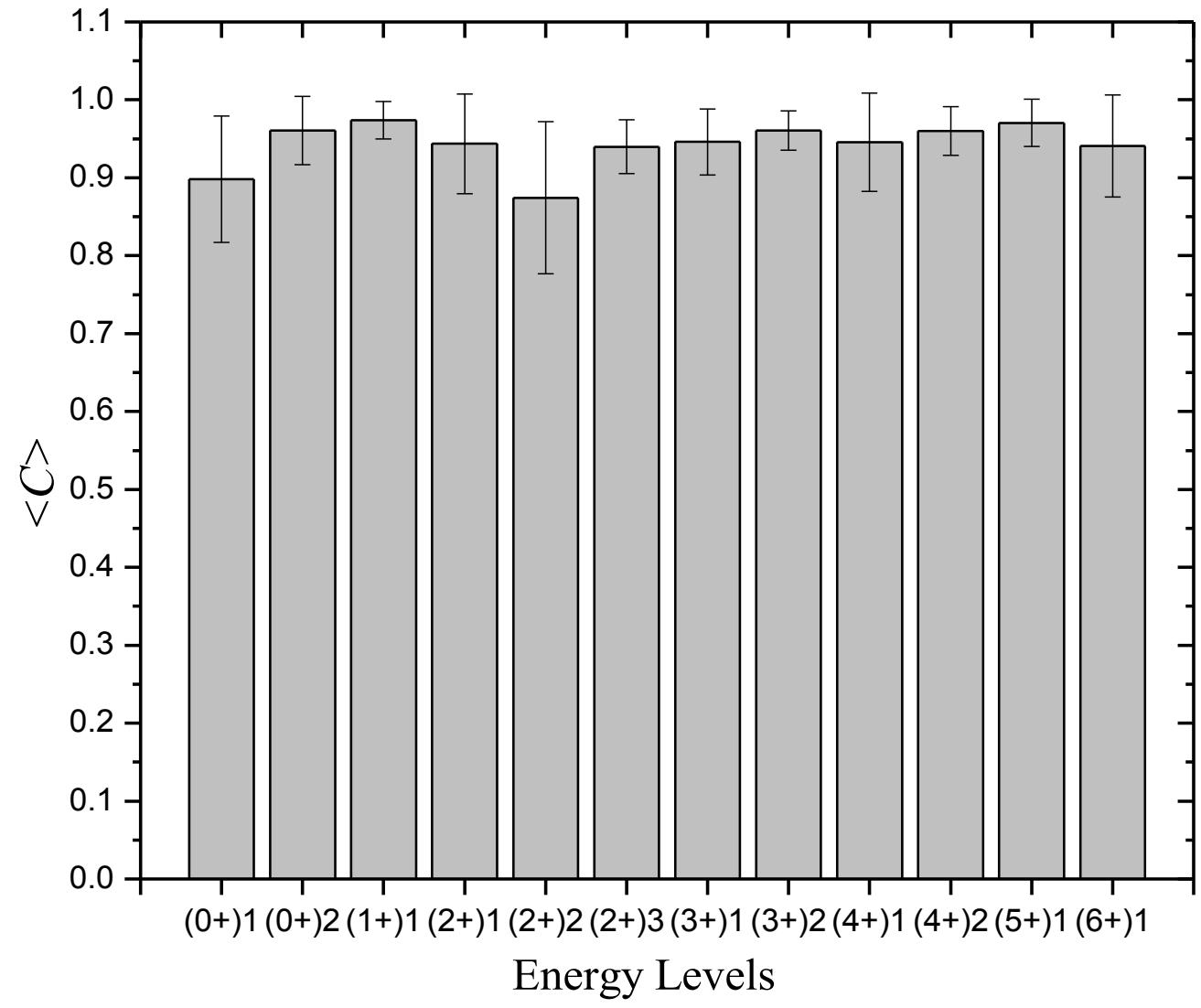

Fig.(2): Mean cross correlation of each positive spin-parity state of *all* nuclei in the investigated range.

## 3-3: Detailed Analysis

In this section, we will follow the systematic analysis outlined by J. Kern *et al*. [28]. Their study was concerned in regions where systematic IBM calculations has been done trying to select nuclei that fitted the *U*(5) symmetry better. Here we perform a similar study on nuclei assumed to satisfy the criteria of the *SU*(3) group.

We qualify best nuclei that exhibit *SU*(3) dynamical symmetry in the range $R_{4/2} \geq 3.25$ by obeying the following conditions,

(i) having $R_{4/2} \geq 3.25$.
(ii) having the $P$-factor value $\geq 5$.
(iii) having states of the same spin in the $\beta$- and $\gamma$-bands nearly degenerate
(iv) satisfying the $SU(3)$ IBM-1 formula for the energy levels, and
(v) satisfying the $SU(3)$ IBM-1 predictions for electromagnetic transitions.

**A: $R_{4/2}$ ratio**

The $R_{4/2}$ ratio systematic enjoys much respect to describe the evolution of nuclear structure. The IBM predicts for an ideal nucleus to have $R_{4/2}$ =3.33 to represent the $SU(3)$ dynamical symmetry. Most deformed nuclei do not reach this fixed value. Instead, we examine the range $R_{4/2} \geq 3.25$ in order to have a sufficient number of levels for subsequent analysis. We consider 38 nuclei according to rules given in section 2. These nuclei are presented in table (1).

**B**: ***P*-factor**:

It is well known that the pairing interaction between like nucleons drives the nucleus towards a spherical shape. It forms the $J = 0^+$ coupling of pairs of identical nucleons that have spherically symmetric wave functions. Deformation and collectivity, on the other hand, arise from configuration mixing which corresponds to a non-uniform distribution of magnetic sub-state occupation and hence, of non-spherical shapes. Configuration mixing itself is largely driven by the valence *p–n* interaction. Hence it is a pairing *p–n* competition that tends to drive the structural evolution of nuclei [29-31]. This idea was used to estimate the locus of collectivity in nuclei. Since the *p–n* interaction strength is, roughly, 200–250 keV, and the pairing interaction is 1 MeV, it takes something like five *p–n* interactions to overcome one pairing interaction. Thus, one expects significant collectivity and the onset of deformation when the $P$-factor [32-34] given in the following formula equals or larger than 5.

$$P = \frac{N_n N_p}{N_n + N_p}, \tag{4}$$

where $N_p$ and $N_n$ are the numbers of valence protons and neutrons, respectively, $N_n N_p$ represents the number of *p–n* interactions and $N_n + N_p$ is the number of pairing interactions. The $P$-factor is given for each nucleus in table (1).

**C: Level degeneracy**

Symmetry infers degeneracy and eigenstates that are degenerate in energy provide a Hilbert space, in which irreducible representations of the symmetry group are valid. One of the important features of the $SU(3)$ dynamical symmetry is the degeneracy of levels having the same spin of the $\beta$- and $\gamma$-bands of the lowest excitation energy $K$=$0^+$, and $K$=$2^+$ of irreducible representations $(\lambda;\mu) = (2N\text{-}4,2)$ respectively [2]. The authors of refs. [35-37] accept the classical picture of the first excited $K$=$0^+$ band as being a $\beta$-vibration. They show that the lowest $K$=$0^+$ band can be of various origins in different nuclei and that several excitation modes compete for the lowest band. Although this difficulty, we will follow the classical interpretation and treat the $\beta$- band as of the first excited $K$=0[+] band. According to the available experimental data, we choose the degeneracy of the first $2^+_\beta$, and the first $2^+_\gamma$ states as an example. We assume that, $2^+_\beta$, and $2^+_\gamma$ states could be considered degenerate if $'\delta = \left| E(2^+_\beta) - E(2^+_\gamma) \right| \le 100$ keV. Table (1) shows the energy level values $E(2^+_\beta)$, $E(2^+_\gamma)$, and $'\delta$ for nuclei under investigation. The following nuclei show this characteristic: $^{158}$Gd, $^{170}$Er, $^{170}$Yb, $^{180}$Hf, $^{182}$W, $^{232}$Th, $^{234}$U, $^{238}$U, $^{238}$Pu, $^{246}$Cm, and $^{248}$Cm.

**D: Level Schemes**

Nuclei claimed to be a member of the *SU*(3) dynamical symmetry must obey the *SU*(3) IBM-1 formula for their energy levels:

$$E_{SU(3)} = E0 - k[\lambda(\lambda+3) + \mu(\mu+3) + \lambda\mu - 2N(2N+3)] + k^{-}L(L+1)\,, \tag{5}$$

where λ, and $\mu$ are the quantum numbers classifying the rotational states, $N$ is the total number of bosons and $L$ is the level spin. The factors $k$, and $k^-$ are adjustable parameters, and listed in table (1). We use two relevant quantities [26] to determine the degree of agreement between observed (experimentally) and calculated (fitted) energy levels using eq. (5) in each nucleus. The first is the average absolute deviation

$$\Delta = \frac{1}{N_L} \sum_i \left| E_i^{\exp} - E_i^{fit} \right|, \tag{6}$$

where $E_i^{\exp}$ and $E_i^{fit}$ are the experimental and best-fit energies in keV of the $i_{\rm th}$ level, while $N_{\rm L}$ is the number of levels. The second is the quality factor defined by

$$Q = \frac{1}{N_L - b} W_i \sum_i \left( E_i^{\exp} - E_i^{fit} \right)^2, \tag{7}$$

where $b$ is the number of the adjustable parameters and $W_i$ =0.01 is a weighing factor chosen to correspond to a uniform uncertainty of 10 keV on the level energies.

We use 16 energy levels to perform the fit procedure using the least-square method: $2^+_1$, $0^+_2$, $2^+_2$, $4^+_1$, $0^+_3$, $2^+_3$, $3^+_1$, $4^+_2$, $6^+_1$, $2^+_4$, $2^+_5$, $4^+_3$, $5^+_1$, $6^+_2$, $7^+_1$, and $8^+_1$. The number of available experimental energy levels (see, table (1)) was variable ranging between 6 in *e.g.* $^{156}$Sm; to 16 in *e.g.* $^{168}$Er. We considered the energy levels of a nucleus to satisfy the *SU*(3) formula if $\Delta \le 100$ keV and $Q \le 150$ keV. The $Q$ and Δ quantities are sensitive to the number of levels involved in fitting procedure. Subsequently, we restrict $N_{\rm L}$ to be not less than 9 levels and at least 5 of them to be members of $\beta$- and/or $\gamma$-bands. The Following nuclei $^{170}$Er, $^{182}$W, $^{232}$Th, and $^{248}$Cm meet this property besides having degenerate levels as given in subsection C. On the other hand, $^{184}$W, $^{232}$U, and $^{256}$Fm nuclei exhibit this feature and lack the degeneracy condition.

**E: Transition Ratio**

For an ideal nucleus having the $SU(3)$ dynamical symmetry, the electromagnetic transitions are very specific and constrained to restricted selection rules depending on the wavefunction of initial and final states of transition. The $\beta$- and $\gamma$-bands form a separate representation from the ground state band, thus the $E2$ transitions from either of these bands to the ground state are forbidden by the $SU(3)$ selection rules. Actually the resultant transition does not vanish, instead it approaches small values. The $B(E2)$ of the $\beta$- and $\gamma$-transition in table (1) are presented in Weisskopf units (w.u.) which allow a rough estimate of the number of nucleons contributing to radiation, and consequently indicate whether the transition is a single particle or collective in nature.

It was shown that [16,38,39] the ratio $R$ between $B(E2{:}2^+_\beta \rightarrow 0^+_g)$ to $B(E2{:}2^+_\gamma \rightarrow 0^+_g)$ comes close to a fixed limiting value of approximately $1/6$ for large boson numbers. We notice from table (1) (taking an arbitrary range of $\pm 0.04$ around the predicted limited value $1/6$) that the following nuclei show signs of this feature $^{160}$Dy, $^{166}$Er, $^{172}$Yb, and $^{178}$Hf. On other hand, The IBM predicts for the $SU(3)$ limit, the transition between $\beta \rightarrow \gamma$ or $\gamma \rightarrow \beta$ to be favored and more collective in nature. Unfortunately we have not been able to find enough experimental data to make fair conclusion. Two nuclei having explicit values of such transition $^{158}\mathrm{Gd}\, 4^+_\beta \rightarrow 2^+_\gamma (12.8\mathrm{w.u.})$, and $^{172}$Yb $2^+_\gamma \rightarrow 2^+_\beta (3.4\mathrm{w.u.})$, $2^+_\gamma \rightarrow 0^+_\beta (2.42\mathrm{w.u.})$.

We note that the $E2$ transitions predicted by $SU(3)$ group introduced a collection of nuclei that do not show degeneracy of levels and/or obeying level schemes as given in subsections C and D. This situation raises the importance of adding separate criteria of the $SU(3)$ limit to give up a reasonable conclusion. We can by a crude estimation classify nuclei in the range under investigation into four main categories. The first category, $^{170}$Er, and $^{182}$W isotopes, show good two signatures of the $SU(3)$: degeneracy, and follow the analytical energy levels formula (eq.(5)), together with a somewhat accepted level of $E2$ transitions. So these isotopes may be the best of the selected nuclei to represent the $SU(3)$ dynamical symmetry in the current study. The second category as the first one exhibit the same two criteria of $SU(3)$ group but the $E2$ transitions does not agree or even missed for the following two nuclei respectively: $^{232}$Th, and $^{248}$Cm. The third category consists of nuclei meet only one property of $SU(3)$ group through current study: $^{158}$Gd, $^{170}$Yb, $^{180}$Hf, $^{234}$U, $^{238}$U, $^{238}$Pu, and $^{246}$Cm meet the degeneracy of levels, while $^{184}$W, $^{232}$U, and $^{256}$Fm having suitable level schemes. The $^{160}$Dy, $^{166}$Er, $^{172}$Yb, and $^{178}$Hf nuclei agree with predictions of $E2$ transitions. The fourth class consists of the rest of nuclei, whether their data are missed to draw a conclusion or even available but do not show criteria of $SU(3)$ limit.

## 4-SUMMARY AND CONCLUSION

IBM-1 is a powerful tool for studying the structure of low-lying excited states of even-even nuclei. It classifies nuclei as belonging to three dynamical symmetries, namely $U(5)$, $SU(3)$, and $O(6)$. Only a limited number of nuclei have been assigned to belong to a given symmetry.

In this study, we focus on deformed nuclei that can exhibit the $SU(3)$ dynamical symmetry. For this purpose, we search for the $SU(3)$ limit signatures in nuclei having $R_{4/2} \geq 3.25$. We start by selecting nuclei having the $P$-factor $\geq 5$ to observe the collective behavior. Next, we turn to examine the collective behavior of that range.

The NNS and level cross-correlation studies validate the regular behavior of selected nuclei. But, we can not use this result alone as an indicator for the existence of the *SU*(3) dynamical symmetry since this regular behavior may come from unidentified conserved or partially conserved quantum numbers not related to the SU(3) symmetry.

We examine in details some main characteristic criteria of the *SU*(3) dynamical symmetry. Each criterion is studied individually. Some nuclei are found to obey each criterion to a good extent. A major problem that faces this study was the insufficiency of the experimental data. Only eighteen nuclei from total thirty-eight were observed to have sufficient data to test their agreement with the *SU*(3) limit predictions. Although this difficulty, we have combined individual evidences together to build a reasonable conclusions.

It is critical to stress that *SU*(3) is very particular case of axially symmetric rotor, and the vast majority of deformed nuclei differ substantially from *SU*(3) limit and require large symmetry breaking to account for their observed properties. This situation, for which only subsets of states obey an exact dynamical symmetry, while other states are mixed, is referred to as partial dynamical symmetry [40]. The mathematical aspects and algorithm for partial dynamical symmetry are presented in Ref. [41].

Table (1): The quality factor $Q$ and the absolute average deviation $\Delta$ for fitted energy levels $N_L$, $P$- factor, the energy levels $E(2^+_\beta)$, $E(2^+_\gamma)$, $B(E2)$ and $\delta$ are shown for nuclei belonging to the interval $R_{4/2} \geq 3.25$.

| **Nucleus** | $N_b$ | $R_{4/2}$ | **P** | $N_L$ | Adjustable parameters of Eq. (5) | | ***SU*(3)** | | $2^+_\beta$ | $2^+_\gamma$ | $\delta$ | $B(E2: 2^+_\beta \to 0^+_g)$ | $B(E2: 2^+_\gamma \to 0^+_g)$ | **R** |
|---|---|---|---|---|---|---|---|---|---|---|---|---|---|---|
| | | | | | $\kappa$ | $\kappa'$ | $\Delta$ | $Q$ | | | | | | |
| $^{152}_{60}$Nd | 10 | 3.263 | 5.00 | 7 | 10.69 | 11.51 | 37.6 | 43.7 | 1251 | -- | -- | -- | -- | -- |
| $^{154}_{62}$Sm | 11 | 3.254 | 5.46 | 15 | 7.25 | 16.68 | 230.9 | 812.5 | 1177 | 1440 | 263 | 0.94 | 3.2 | 0.29 |
| $^{156}_{62}$Sm | 12 | 3.290 | 6.00 | 6 | 8.89 | 12.94 | 59.2 | 112.9 | -- | -- | -- | -- | -- | -- |
| $^{158}_{64}$Gd | 13 | 3.288 | 6.46 | 15 | 5.93 | 14.3 | 162.2 | 384.5 | 1259 | 1187 | 72 | 0.31 | 3.4 | 0.09 |
| $^{160}_{66}$Dy | 14 | 3.270 | 6.86 | 16 | 5.3 | 14 | 110 | 395.2 | 1349 | 966 | 383 | 0.65 | 4.5 | 0.14 |
| $^{160}_{64}$Gd | 14 | 3.302 | 7.00 | 15 | 5.16 | 12.32 | 121.1 | 441.9 | 1377 | 988 | 389 | -- | 3.8 | -- |
| $^{162}_{64}$Gd | 15 | 3.291 | 7.47 | 6 | 4.54 | 11.55 | 3.5 | 0.3 | 1492 | 864 | 628 | -- | -- | -- |
| $^{162}_{66}$Dy | 15 | 3.294 | 7.47 | 15 | 5.38 | 11.23 | 110.2 | 256.4 | (1728) | 888 | 840 | -- | 4.6 | -- |
| $^{164}_{68}$Er | 14 | 3.276 | 7.00 | 15 | 4.78 | 14.12 | 127.8 | 428.7 | 1314 | 860 | 454 | 0.36 | 5.2 | 0.07 |
| $^{164}_{66}$Dy | 16 | 3.302 | 8.00 | 12 | 5.19 | 10.16 | 178 | 511 | -- | 761 | -- | -- | 4 | -- |
| $^{166}_{68}$Er | 15 | 3.289 | 7.47 | 15 | 5.7 | 9.94 | 236.4 | 870.6 | 1528 | 786 | 742 | 0.66 | 5.17 | 0.13 |
| $^{166}_{66}$Dy | 17 | 3.310 | 8.47 | 9 | 4.69 | 10.03 | 117.6 | 230.9 | 1208 | 857 | 351 | -- | -- | -- |
| $^{168}_{70}$Yb | 14 | 3.266 | 7.00 | 15 | 4.84 | 14.84 | 151.8 | 455.9 | 1233 | 983 | 250 | 1.8 | 4.6 | 0.39 |
| $^{168}_{68}$Er | 16 | 3.309 | 7.88 | 16 | 4.55 | 11.87 | 140.7 | 369.9 | 1276 | 821 | 455 | -- | 4.8 | -- |
| $^{170}_{68}$Er | 17 | 3.310 | 8.24 | 16 | 3.66 | 14 | 86.8 | 109.5 | 960 | 934 | 26 | 0.28 | 3.68 | 0.08 |
| $^{170}_{70}$Yb | 15 | 3.293 | 6.00 | 16 | 3.53 | 16.02 | 171.8 | 434.7 | 1138 | 1145 | 7 | 1.08 | 2.7 | 0.4 |
| $^{172}_{70}$Yb | 16 | 3.305 | 7.50 | 16 | 4.63 | 14.1 | 133.7 | 365.9 | 1117 | 1465 | 348 | 0.24 | 1.33 | 0.18 |
| $^{174}_{72}$Hf | 15 | 3.269 | 6.67 | 14 | 4.54 | 15.07 | 132.2 | 392.7 | 900 | 1226 | 326 | 2.1 | 4.8 | 0.44 |
| $^{176}_{72}$Hf | 16 | 3.284 | 6.88 | 15 | 4.77 | 17.66 | 231 | 785.4 | 1226 | 1341 | 115 | 0.98 | 3.9 | 0.25 |
| $^{178}_{72}$Hf | 15 | 3.291 | 6.67 | 15 | 5.11 | 17.04 | 177.9 | 470.6 | 1276 | 1174 | 102 | 0.72 | 3.9 | 0.18 |
| $^{178}_{70}$Yb | 15 | 3.309 | 7.20 | 7 | 7.25 | 13.78 | 36.1 | 41.2 | 1404 | 1221 | 183 | -- | -- | -- |
| $^{180}_{74}$W | 14 | 3.260 | 5.71 | 10 | 7.15 | 15.2 | 97.5 | 218.2 | -- | 1117 | -- | -- | -- | -- |
| $^{180}_{72}$Hf | 14 | 3.307 | 6.43 | 14 | 5 | 17.15 | 203 | 594.1 | 1183 | 1199 | 16 | -- | 3.8 | -- |
| $^{182}_{74}$W | 13 | 3.291 | 5.54 | 14 | 7 | 16.9 | 76.8 | 100.6 | 1257 | 1221 | 36 | 0.91 | 3.4 | 0.27 |
| $^{184}_{74}$W | 12 | 3.274 | 5.33 | 15 | 5.51 | 17.72 | 76.2 | 148.5 | 1121 | 903 | 218 | 0.21 | 4.41 | 0.05 |
| $^{230}_{90}$Th | 11 | 3.273 | 5.09 | 15 | 5.39 | 8.34 | 86.2 | 223.3 | 677 | 781 | 104 | 1.1 | 3 | 0.37 |
| $^{232}_{90}$Th | 12 | 3.284 | 5.33 | 16 | 4.62 | 8.83 | 69.2 | 76.9 | 774 | 785 | 11 | 2.8 | 2.9 | 0.97 |
| $^{232}_{92}$U | 12 | 3.291 | 5.39 | 13 | 4.67 | 8.26 | 79.9 | 144.6 | 734 | 866 | 132 | -- | -- | -- |
| $^{234}_{92}$U | 13 | 3.295 | 6.15 | 14 | 4.51 | 5.59 | 134.2 | 255.6 | 851 | 926 | 75 | < 1.3 | 2.9 | <0.49 |
| $^{238}_{92}$U | 15 | 3.305 | 6.67 | 14 | 3.9 | 9.39 | 188 | 504.2 | 966 | 1060 | 94 | 0.38 | 3.04 | 0.13 |
| $^{238}_{94}$Pu | 15 | 3.311 | 7.20 | 12 | 4.2 | 8.23 | 157.3 | 394.9 | 983 | 1028 | 45 | 3.9 | -- | -- |
| $^{240}_{94}$Pu | 16 | 3.309 | 7.50 | 13 | 3.82 | 8.25 | 155.1 | 396.6 | 900 | 1137 | 237 | -- | -- | -- |
| $^{242}_{94}$Pu | 17 | 3.307 | 7.77 | 11 | 3.61 | 7.58 | 128.1 | 358.4 | 992 | 1102 | 110 | -- | -- | -- |
| $^{244}_{96}$Cm | 18 | 3.313 | 8.56 | 8 | 4.52 | 6.15 | 57.2 | 75.7 | -- | -- | -- | -- | -- | -- |
| $^{246}_{96}$Cm | 19 | 3.314 | 8.84 | 12 | 3.75 | 8.08 | 189.4 | 601.8 | 1210 | 1124 | 86 | -- | -- | -- |
| $^{248}_{96}$Cm | 20 | 3.309 | 9.10 | 9 | 4.5 | 6.99 | 22.2 | 10.6 | 1126 | 1049 | 77 | -- | -- | -- |
| $^{250}_{98}$Cf | 21 | 3.321 | 9.91 | 12 | 3.15 | 7.89 | 181.8 | 576.5 | 1189 | 1031 | 158 | -- | 2.3 | -- |
| $^{256}_{100}$Fm | 24 | 3.310 | 11.2 | 11 | 2.43 | 7.45 | 62.5 | 132.1 | -- | 682* | -- | -- | -- | -- |

#brackets represent uncertainty in selected value.
#"*" indicates that this value taken from the Table of Isotopes, 8$^{th}$ edition.